\documentclass[twocolumn,showpacs,preprintnumbers,amsmath,amssymb]{revtex4}
\usepackage{amsmath}
\usepackage{graphicx}
\usepackage{color}

\begin{document}
\title{Mechanical Manipulations on Electronic Transport of Graphene Nanoribbons}

\author{Jing Wang$^1$}
\author{Guiping Zhang$^{1}$}\email{bugubird_zhang@hotmail.com}
\author{Fei Ye$^2$}
\author{Xiaoqun Wang$^{1,3,4,5}$}\email{xiaoqunwang@ruc.edu.cn}
\affiliation{$^1$Department of Physics, Renmin University of China,
  Beijing 100872, China}
\affiliation{$^2$Department of Physics,  South University of Science
  and Technology of China, Shenzhen 518055, China}
\affiliation{$^3$Beijing Laboratory of Optoelectronics Functional Materials and Micronano Device, Renmin University of China, Beijing 100872, China}
\affiliation{$^4$Department of Physics and Astronomy, Shanghai Jiao Tong University,
  Shanghai 200240, China}
\affiliation{$^5$ Collaborative Innovation Center of Advanced Microstructures, Nanjing 210093, China}
\date{\today}

\begin{abstract}
  We study the effects of uniaxial strains on the transport properties
  of the graphene nanoribbons(GNRs) connected with two metallic leads in
  heterojunctions, using the transfer matrix method. Two typical GNRs
  with zigzag and armchair boundaries are considered, and the tension
  is applied either parallel or perpendicular to the ribbon axis. It
  turns out that the electron-hole symmetry is missing in the gate voltage
  dependence of the conductance data of the armchair GNRs, while it
  persists in the zigzag ribbons under any strains. For an armchair GNR
  with a vertical tension applied, a sharp drop of conductance is found
  near the critical value of the strain inducing a quantum phase transition,
  which allows to determine the critical strain accurately via measuring
  the conductance. In the zigzag ribbon, there exists a range of gate
  voltage around zero, where the conductance is insensitive to the small
  horizontal strains. The band structures and low-energy properties are
  calculated to elucidate the mechanism on the strain effects in GNRs.
  We expect that our results can be useful in developing graphene-based
  strain sensors.
\end{abstract}
\maketitle

\section{Introduction}
Graphene has attracted widespread interests both theoretically and
experimentally since its discovery in 2004, because of its unique
properties and promising
applications\cite{graphene1,graphene2,graphene3-e}. Recently, there are
many studies on its mechanical deformation\cite{strain-e5,strain-e3} and
the corresponding effects on the Raman spectroscopy\cite{strain-e6},
since the strain is inevitable for the fabrication of graphene on
substrate.  Unlike the conventional materials, graphene has a tough
mechanical property and could sustain elastic deformation up to
$15\sim20\%$\cite{elastic1,elastic2}.  There are two typical ways to
control the strain in graphene samples\cite{strain-e6,elastic1}. One way
is using the subtrate with an array of holes with diameter ranging from
$1\mu$m to 1.5$\mu$m. When a free-standing monolayer graphene is
transferred onto it, a nonlinear strain-stress relation is observed by
nanoindentation in the atomic force microscope\cite{elastic1}, which has
been verified theoretically\cite{elastic2}.  The other one is by
exerting tension on the subtrate to control the strain on the
graphene\cite{strain-e6,strain-e1,strain-e2}. Graphene ripples on
polydimethylsiloxane(PDMS) substrate can afford a reversible structural
deformation under tensile strains as large as
$20\sim30\%$\cite{strain-e4}.

At atomic scale, the C-C bond length is changed by the strain, so are
the hopping integrals and the band structure of graphene. To open a band
gap in graphene, one requires a uniaxial strain in excess of 20\%, which is beyond the range of elastic deformation\cite{strain-TB}. In contrast, a much smaller uniaxial strain can control (close or open) the band gap in the narrow armchair graphene nanoribbons(AGNRs), while the zigzag graphene nanoribbons(ZGNRs) are quite
robust against gap opening for small strain
\cite{strain-band
  structure1, strain-band structure2, strain-band structure3}.

In recent years, the field of graphene-based strain sensors develops rapidly, since it is feasible to mediate electronic properties of graphene by applying tensions. In a sample of graphene from chemical vapor deposition, the resistance remains around 7.5$K\Omega$ under the
strain less than 2.47$\%$ applied along the electronic transport direction, while increases rapidly to 25$K\Omega$ under 5$\%$ strain\cite{strain-e1}. This is because that the ripples in graphene do not disappear until the strain exceeds 2.47$\%$. The strain dependent transport
properties enable graphene to have potential applications in the fields of the displays, robotics, fatigue detection, body monitoring, and so forth. For instance, the graphene-based strain sensors on the transparent gloves can measure the magnitudes and
directions of the principal strains on the glove induced by the motion of fingers\cite{strain-e2}.

 {The previous theoretical investigations on the transport
  properties of the strained graphene
  nanoribbons\cite{strain-transport1,strain-transport2,strain-transport3,strain-transport4,strain1,strain2,strain3}
  mostly deal with the small-scale GNRs with a width of about several
  nanometers using homojunction contacts,
  while one may encounter more complicated situations in practice,
  e.g., heterojunction contacts and wider GNRs in the fabrication of the
  GNR-based nanodevices.} In this paper, we utilize a transfer matrix
method\cite{tm1} to study the transport properties of both narrow and
wide GNRs under the strain, which are in particular connected to two
metallic leads with heterojunctions. The width of graphene can reach the
order of microns by means of the transfer matrix method\cite{tm1}.  A
tight binding model is taken to describe the low energy physics for both
the $\pi$-electrons of graphene and metallic electrons in two leads. The
effects of strains on the hoping integral of C-C bonds in graphene are
elucidated in Sec.II. The band structures of AGNRs and ZGNRs for various
sizes under different strains are presented in Sec.III. In Sec. IV, we
show the effects of strains on the transport properties of both AGNRs
and ZGNRs. The edge effects are discussed in Sec. V. We note that strain
only affects the band structure
of graphene and the electronic transport of strained graphene we present
embodies the combined effects of strains and the heterojunctions
composed of graphene and metallic contacts.

\section{Model and method}
\subsection{Strained Graphenes}
\label{sec:strained-graphene}
\begin{figure}[htb]
  \centering
  \includegraphics[width=8cm]{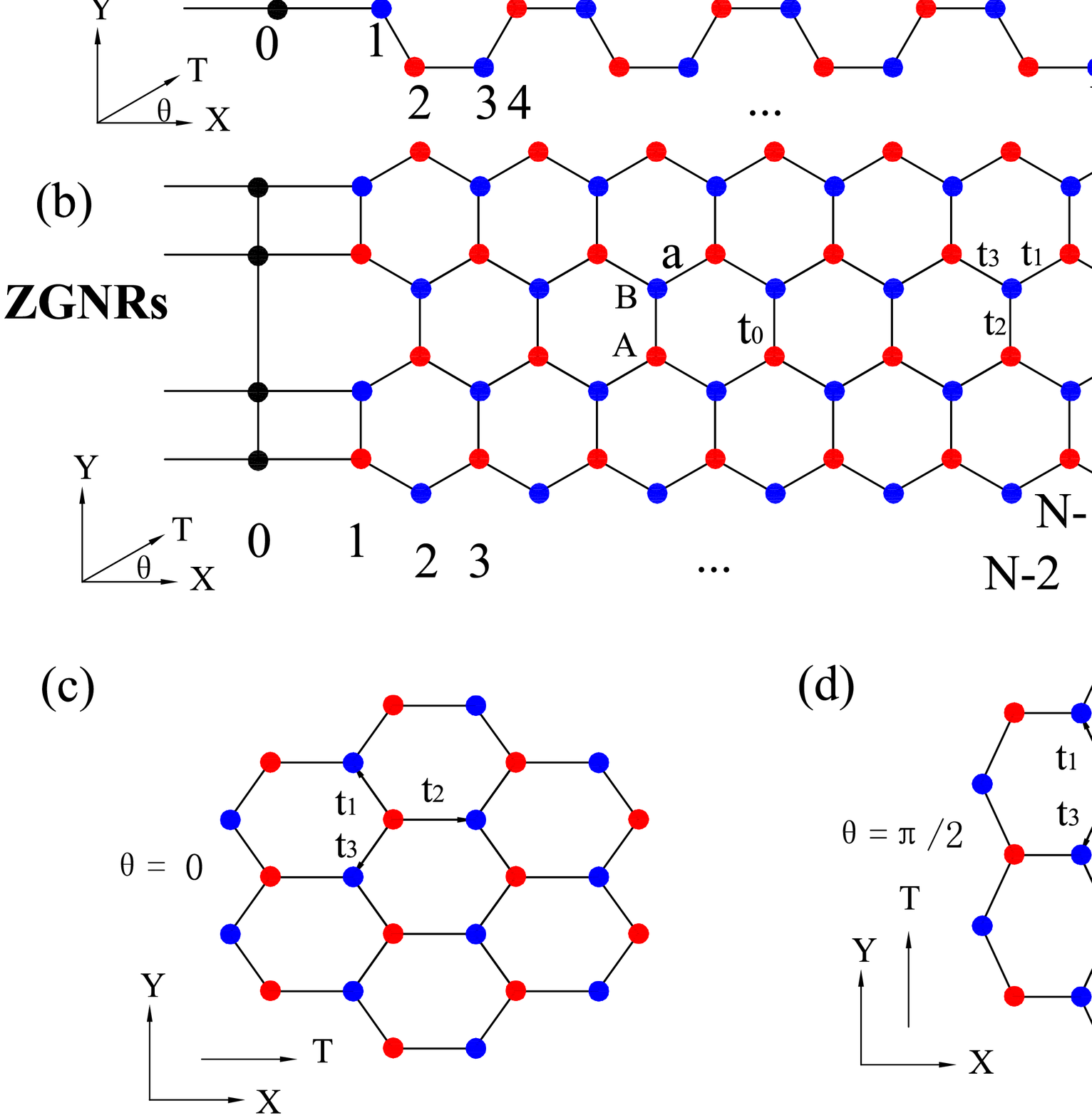}
  \caption{\label{QW_GRs}(Color online.) The schematic illustration of
    AGNRs (a) and ZGNRs (b), connected to two semi-infinite quantum
    wires.  There are $N$ and $M$ carbon atoms in $x$ and $y$
    directions, respectively. (c) AGNRs with the tension
      along $x$-axis, i.e. $\theta=0$, and (d) along $y$-axis with
      $\theta=\pi/2$.}
\end{figure}

To investigate the influence of uniaxial strain on the electronic transport properties of GNRs, we connect it with semi-infinite quantum wires, which are characterized by the square lattices, as illustrated in Fig. \ref{QW_GRs}. Each interface between the GNR and a lead is a heterojunction. For AGNRs in Fig.~\ref{QW_GRs}(a), the interface is a ring consisting of five atoms, which eventually breaks the electron-hole(e-h) symmetry of the system, while for ZGNRs in Fig.~\ref{QW_GRs}(b), each ring at two interfaces contains four or six atoms which retains the e-h symmetry since the tight binding model involves only the nearest neighbor hopping in this paper. This is revealed by the dependence of conductance on gate voltages as shown
later. The uniaxial tension is only applied to the GNRs leading to the deformation of C-C bonds in an anisotropic way.

The strain-stress relation for graphene is given in
Ref. \cite{strain-TB}. We quote those relevant results here for our
further discussions. The tension is applied along the direction
$\cos\theta \vec{e}_x + \sin\theta \vec{e}_y$, and the corresponding
tensile strains parallel and perpendicular to this direction are $S$ and
$-\nu S$, respectively, with the Poisson's ratio
$\nu=0.165$\cite{Poisson-ratio}.  In the lattice coordinate system the
strain tensor reads
\begin{equation}
\label{eq:1}
\boldsymbol{\epsilon} = S \left(
  \begin{array}{ccc}
    \cos^2\theta-\nu\sin^2\theta & (1+\nu)\cos\theta\sin\theta \\
     (1+\nu)\cos\theta\sin\theta & \sin^2\theta-\nu\cos^2\theta
  \end{array}
\right).
\end{equation}
For any vector $\vec{l}_0$ in the undeformed graphene plane, it is
straightforward to obtain its deformed counterpart to the leading order
by the transformation
\begin{eqnarray}
\label{eq:2}
\vec{l}=(1+\boldsymbol{\epsilon})\vec{l}_0.
\end{eqnarray}

The hopping amplitude $t_{i}$ with $i=1,2,3$ as defined in
Fig.~\ref{QW_GRs} is determined by the corresponding bond length
$\delta_{i}$ via the following formula\cite{strain-TB}
\begin{equation}
\label{eq:3}
t_{i}=t_0e^{-3.37(\frac{\delta_i}{a}-1)},
\end{equation}
with $t_0=2.6eV$ and $a=1.42\AA$ for the undeformed graphene. The bond
length $\delta_i$ under the strain can be calculated by Eq.~\eqref{eq:1}
and Eq.~\eqref{eq:2}.  Without loss of generality, we focus on two cases
with $\theta=0$ and $\pi/2$ as follows.
\begin{itemize}
\item For $\theta=0$ shown in Fig.~\ref{QW_GRs}(c),
\begin{equation*}
\delta_1=\delta_3= (1+\frac{1}{4}S-\frac{3}{4}\nu S)a, \hspace{0.3cm} \delta_2=(1+S)a.
\end{equation*}
All the three bond lengths increase as $S$ increases, and the $t_{i}$'s
subsequently decrease for all $i=1,2,3$. However, $\delta_2$ increases
faster than $\delta_1$ and $\delta_3$.  Therefore, we have $t_1=t_3>t_2$
as long as $S>0$.
\item For $\theta=\pi/2$ shown in Fig.~\ref{QW_GRs}(d),
\begin{equation*}
\delta_1=\delta_3=(1+\frac{3}{4}S-\frac{1}{4}\nu S)a, \hspace{0.3cm}\delta_2=(1-\nu S)a.
\end{equation*}
In this case, we also have $\delta_1=\delta_3$ and $t_1=t_3$. As $S$
increases, $\delta_{1,3}$ increase, while $\delta_2$ decreases. It turns
out that $t_1$ and $t_3$ decrease and $t_2$ increases with increasing
$S>0$.
\end{itemize}

For ZGNRs, the lattice coordinate system is rotated by $\pi/2$ from that
of AGNRs.  Therefore, the strain effect on the hopping amplitudes for
ZGNRs with $\theta=0$ (or $\pi/2$) is identical to that for AGNRs with
$\theta=\pi/2$ (or $0$). The hopping amplitudes $t_i$'s as functions of
$S$ are plotted for AGNRs in Fig.~\ref{band_AGR}(a), in the unit of
$t_0$, which is set as one in the following discussions.

\subsection{Tight-binding Model and Transfer Matrix Method}

The $\pi$-electrons of carbon atoms are responsible for the low energy physics of graphene which can be described by the tight binding model on the honeycomb lattice
\begin{equation}
\label{eq:4}
\hat{H}=\sum_{\langle ij,i^{'}j^{'}\rangle}t_{ij,i^{'}j^{'}} \hat{c}^{\dag}_{ij}\hat{c}_{i^{'}j^{'}}
+V_{g}\sum_{ij}\hat{c}^{\dag}_{ij}\hat{c}_{ij},
\end{equation}
where a pair of integers $ij$ indicates the lattice position
$\vec{R}_{ij}=x_i\vec{e}_x+y_j\vec{e}_y$, and $\hat{c}_{ij}$
($\hat{c}^{\dag}_{ij}$) is the corresponding electron
annihilation(creation) operator. The summation is over the nearest
neighbors indicated by $\langle \cdots\rangle$, and $t_{ij,i^{'}j^{'}}$
is the hopping amplitude which takes the value of $t_1$, $t_2$ or $t_3$
depending on the relative position
$\vec{R}_{i^{'}j^{'}}-\vec{R}_{ij}$. The spin indices of electrons are
  omitted simply for convenience.   { $V_{g}$ is the gate
  voltage which is applied only to the GNRs, not on the leads. In our
  simulation, we consider a simplified case with $V_g$ changing abruptly
  at the interfaces between the GNR and the leads. In fact, this
  simplification is reasonable for small $V_g$. For large $V_g$, there
  may exist a junction between the leads and GNR with finite width of
  several atoms. This situation would not be considered here, since
  it only incurs further unnecessary complexities as far
  as the strain effects are concerned.}

The left and right electrodes are also described by the Hamiltonian in
Eq. \eqref{eq:4} with $V_{g}=0$, but the lattice vectors $\vec{R}_{ij}$
 {describe a rectangular lattice instead of the hexagonal
  one. All the hopping integrals in the leads are fixed as $t_0$,
  despite the vertical lattice constants may not be uniform in the leads
  connected to ZGNRs as shown in Fig.~\ref{QW_GRs}b. We also assume the
  leads are unaffected by the strain in our numerical simulation. This
  idealized setup mimics a normal-metal/GNR heterojunction, by which we
  shall demonstrate the strain effects on the transport through GNRs. }

The single-particle eigenstate with energy $E$ can be expressed as
$\hat{\psi}^{\dagger}(E) = \sum_{ij}\alpha_{ij}\hat{c}^{\dagger}_{ij}$,
which satisfies $[\hat{\psi}(E),\hat{H}]=E\hat{\psi}(E)$, leading to
\begin{equation}
(E-V_g)\alpha_{ij}=\sum_{\langle ij,i^{'}j^{'}\rangle}t_{ij,i^{'}j^{'}}\alpha_{i^{'}j^{'}}.
\end{equation}
The wavefunctions of the electrodes can be represented in terms of two
numbers $k_{x}$ and $k_{y}$, where $k_x$ describes the plane wave
traveling along the $x$ direction and $k_y$ is quantized as $k_{y,n}=
n\pi/(M+1)$ with $n=1,2,\cdots,M$ due to the open boundary condition
imposed in the $y$ direction, to characterize different channels. The
corresponding eigenenergy reads
\begin{eqnarray}
\label{eq:5}
E=2t_0(\cos k_{y,n}+\cos k_{x,n}),
\end{eqnarray}
which determines the wave number $k_{x,n}$ in $n$-th channel for the
given Fermi energy $E$.  {Note that, since the hopping
  amplitudes of all the bonds are given, the lattice constant is not
  needed anymore and one can simply use dimensionaless wave numbers
  $k_{x,n}$ and $k_{y,n}$ to label the quantum states.}

If we assume the electrons are incident from the left, the wavefunctions in the left and right electrodes can be written as\cite{tm1,tm4}
\begin{align}
\label{eq:6}
\alpha_{ij}^{L}=&\sum_{n^{'}}(\delta_{n^{'}n}e^{ik_{x,n^{'}}x_i}
+r_{n^{'}n}e^{-ik_{x,n^{'}}x_i})\sin(k_{y,n^{'}}y_j), \nonumber\\
\alpha_{ij}^{R}=&\sum_{n^{'}}t_{n^{'}n}e^{ik_{x,n^{'}}x_i}\sin(k_{y,n^{'}}y_j),
\end{align}
where $t_{n^{'}n}$ and $r_{n^{'}n}$ are the transmission and reflection amplitudes from $n$-th to $n^{'}$-th channel, respectively. Current conservation requires
$\sum_{n^{'}}\eta_{n,n^{'}}[|t_{n,n^{'}}|^2+|r_{n,n^{'}}|^2]=1$ for each $n$ with $\eta_{n,n^{'}}\equiv|\sin (k_{x,n^{'}})|/{|\sin (k_{x,n})|}$. In order to calculate the transmission coefficients $t_{n,n^{'}}$, we adopt the transfer matrix method developed in
Ref. {\onlinecite{tm1}}, and then we can designate $\alpha_j$ for the $M$ coefficients with column index $j$, which satisfies the matrix equation
\begin{equation}
\left(
  \begin{array}{c}
    \alpha_j\\
    \alpha_{j+1}\\
  \end{array}
\right)=\chi_j\left(
\begin{array}{c}
    \alpha_{j-1}\\
    \alpha_j\\
  \end{array}
\right),
\end{equation}
where $\chi_{j}$ is the $2M\times 2M$ transfer matrix as a function of Fermi energy $E$, gate voltage $V_{g}$ and the hopping amplitudes $t_i$'s.  By acting the transfer matrices consecutively, the coefficients in the left and right interfaces are connected in the following form
\begin{equation}
\label{eq:7}
\left(
  \begin{array}{c}
    \alpha_N\\
    \alpha_{N+1}\\
  \end{array}
\right)=\chi_N\chi_{N-1}\ldots
\chi_2\chi_1\left(
\begin{array}{c}
    \alpha_0\\
    \alpha_1\\
  \end{array}
\right).
\end{equation}
Combining Eq.~\eqref{eq:6} and Eq.~\eqref{eq:7} one can obtain the transmission and reflection coefficients $t_{n,n^{'}}$ and $r_{n,n^{'}}$. In order to investigate the transport properties of the large scale GNRs, we actually utilize the renormalized transfer matrix method as described
in Ref. \onlinecite{tm3}. It is straightforward to calculate the conductance by employing Landauer-B\"{u}ttiker formula
\begin{equation}
\label{eq:8}
G=\frac{2e^{2}}{h}\sum_{n,n^{'}=1}^{M}\eta_{n,n^{'}}|t_{n,n^{'}}|^{2},
\end{equation}
where the factor 2 is a consequence of the spin degeneracy.

\section{Band structure of Strained Graphene Nanoribbons}
In this section, we study the band structure and low energy excitations
of GNRs under strains by solving the tight binding model
Eq.~\eqref{eq:4} with the hopping amplitudes given by
Eq.~\eqref{eq:3}. For convenience, we impose the periodic boundary
condition along $x$-axis and the open boundary condition along $y$-axis,
and the energy $E$ is taken in unit of $t_{0}$ in the following
discussions. We also assume the horizontal lattice spacing to be unit so
that $k_x$ is always in the interval $[0,2\pi)$, although it changes as
the tension is applied.


\subsection{Strained AGNRs}
\label{sec:strained-agnrs}
\begin{figure}[htb]
\includegraphics[width=8.8cm]{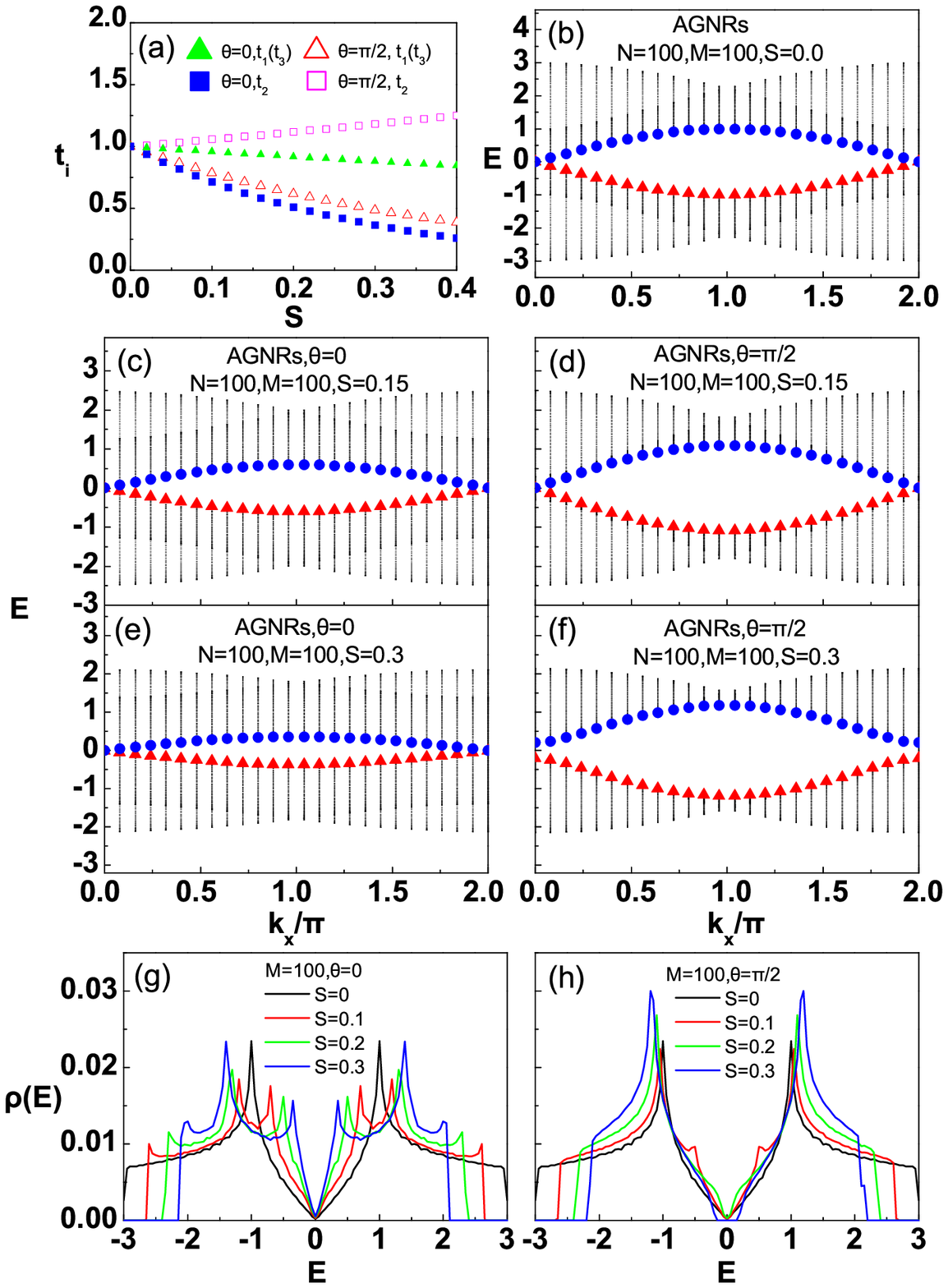}
\caption{\label{band_AGR}(Color online.) (a) shows the
    strain dependence of the three hopping amplitudes.  (b)-(f) are the
    band structures of AGRNs with $N=M=100$ under various strains.
  (g) and (h) are the density of states $\rho(E)$ of AGNRs with
  $M=100$ and $N=1600$ for the tensions along $x$ and $y$ axis,
  respectively.}
\end{figure}

The spectra of AGNRs are plotted in Figs. \ref{band_AGR}(b)-(f) as
functions of $k_x$ with $N=100$ and $M=100$ for different strains. The
unstrained data is given in Fig.\ref{band_AGR}(b) which is precisely
gapless at $k_x=0$.  In the presence of strains, the spectrum changes
upon the direction of the applied tension. When the tension is applied
horizontally to AGNRs, i.e. $\theta=0$, the spectra in
Figs. \ref{band_AGR}(c) and (e) are similar to the unstrained case,
except that the uniaxial strain may open a small gap at $k_x=0$. This
gap is proportional to $M^{-1}$ and becomes almost invisible for
$M=100$, which results from the combined effect of the finite ribbon
widths and the strains.  When the tension is applied vertically,
i.e. $\theta=\pi/2$, the spectrum in Fig.~\ref{band_AGR}(d) with strain
$S=0.15$ shows a tiny gap, which is also proportional to $M^{-1}$ with
the same origin of that for $\theta=0$. For $S=0.3$, another type of gap
opens at $k_x=0$ as shown in Fig.~\ref{band_AGR}(f), which is induced
entirely by the strain \cite{strain-TB} and can survive the
thermodynamic limit unlike the previous gaps. In fact there is a
critical strain $S$ separating the two different gaps as to be discussed
in details later.

Figures \ref{band_AGR}(g) and (h) show the density of states (DOS)
$\rho(E)$ for AGNRs under different strains for $\theta=0$ and $\pi/2$,
respectively. The band width is $D=2t_1+t_2$ plus a negligible
dependence on the ribbon width, which obviously shrinks as $S$ increases
for both $\theta=0$ and $\theta=\pi/2$, although $t_1$($t_3=t_1$) and
$t_2$ behave very differently as functions of $S$ in
Fig.~\ref{band_AGR}(a). When $\theta=0$, $t_1$ and $t_2$ are decreasing
function of $S$ so is the band width. When the strain increases for
$\theta=\pi/2$, $t_2$ increases, but it is the decreasing $t_1$ that
dominates the strain dependence of the band width. Besides the shrinking
band width, there are no other common features in the DOS for both cases
with $\theta=0$ and $\theta=\pi/2$.  For the unstrained ribbons, there
are two peaks of DOS located at $\pm t_0$. Each of them splits into
double peaks if the tension is applied horizontally, which locate at
$\pm t_2$ and $\pm (2t_1-t_2)$ as seen in Fig.~\ref{band_AGR}(g). When
the tension is applied vertically, the peaks at $E=\pm t_2$ in
Fig.~\ref{band_AGR}(h) move slightly outwards instead of
splitting. However no peaks are observed at $E=\pm(2t_1-t_2)$, except
two shoulders emerging at $\pm0.5t_0$ for $S=0.1$ as a remnant of the
peaks. For other strains in Fig.~\ref{band_AGR}(h), even the shoulders
can not be seen.  When the strain $S=0.3$, the DOS vanishes in the
energy range $[-0.205,0.205]$ implying a gap of $0.409(t_0)$ opens.

\begin{figure}[htbp]
\includegraphics[width=8.7cm]{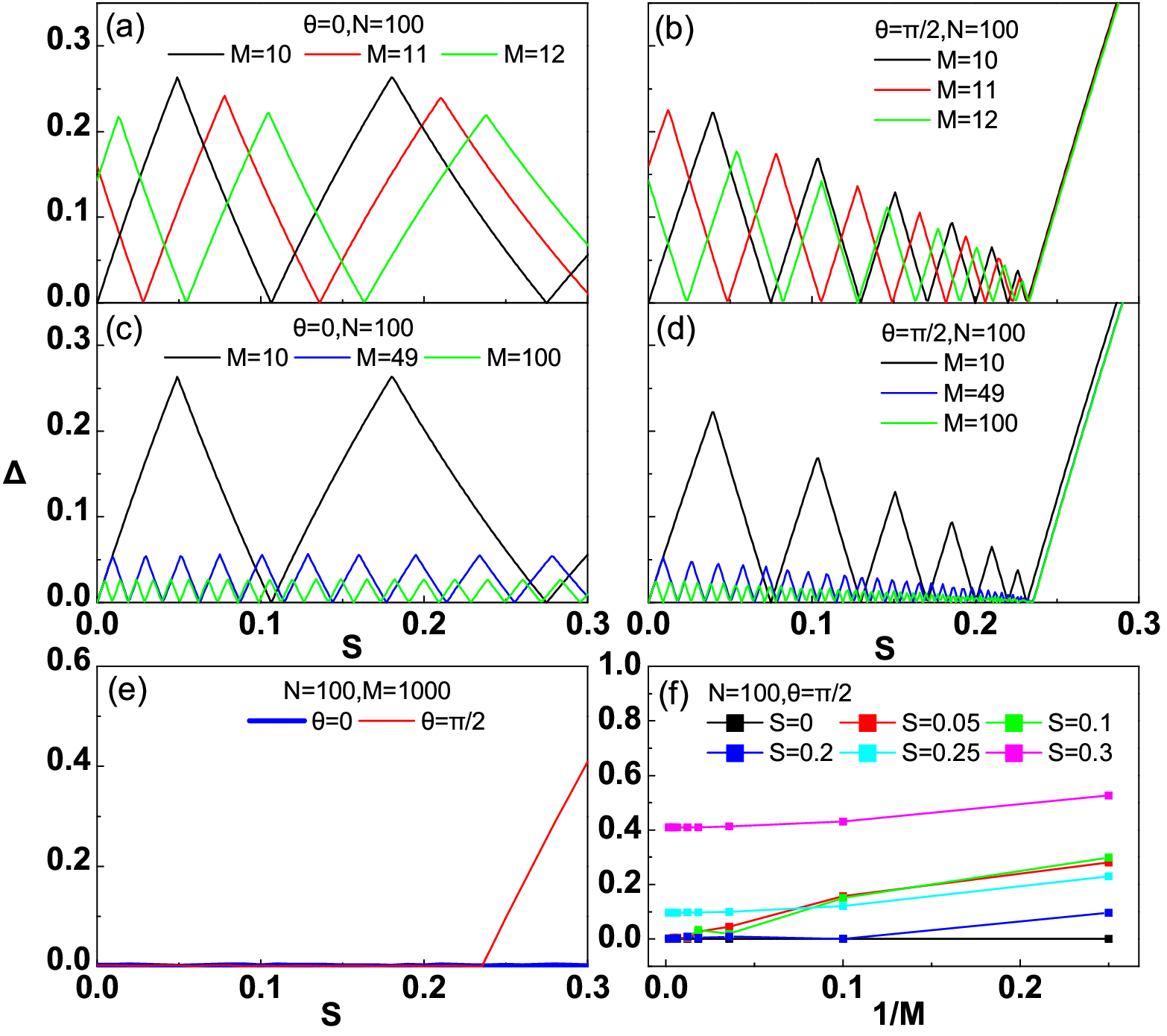}
\caption{\label{gap_AGR}(Color online) Band gaps of
    AGNRs as functions of strains with various sizes. The tension is
    applied horizontally in (a) and (c), and vertically in (b) and
    (d). A very large ribbon width $M=1000$ is taken in (e) for both
    $\theta=0$ and $\pi/2$, where the finite size effect is too small to
    be observable. (f) shows the extrapolation of the band gap to the
    infinite width limit under various strains in the case of
    $\theta=\pi/2$.}
\end{figure}

Now we turn to the dependence of energy gaps at $k_x=0$ on the ribbon
widths and strains. For the unstrained AGNRs, the gap is zero for
$\text{mod}(M,3)=1$ and inversely proportional to $M$ for
$\text{mod}(M,3)=0,2$,  {which coincides with} previous
studies using the first-principle calculation \cite{Yao-AGNR-bandgap}
and the tight binding model\cite{band structure-AGNR,tm1}.
This feature is manifested in Figs.~\ref{gap_AGR}(a)-(d) for $M=10$,
$11$, $12$, $49$ and $100$ with a fixed ribbon length $N=100$.

When the tension is applied horizontally with different ribbon widths,
the band gaps oscillate with the strains in the similar zigzag patterns,
but with different ``phases'' according to different values of
$\text{mod}(M,3)$ as shown in Fig.~\ref{gap_AGR}(a). The oscillatory
amplitude is inversely proportional to $M$ and barely changes with $S$,
and the oscillatory frequency increases with $M$, but decreases with
$S$ as shown in Fig.~\ref{gap_AGR}(c).

For $\theta=\pi/2$ as shown in Figs.~\ref{gap_AGR}(b) and (d), the band
gap behaves significantly different from that for $\theta=0$. The
oscillation only happens for $S<S_c$. In this region, the oscillatory
amplitude decreases with both $M$ and $S$, while the frequency increases
with both $M$ and $S$. The band gaps for both $\theta=0$ and
$\theta=\pi/2$ result from the combined effects of the strains and
the finite ribbon widths. Actually they are almost invisible in
Fig.~\ref{gap_AGR}(e) for $M$ as large as $1000$. The essential
difference occurs for $S>S_c$, where a gap opens for $\theta=\pi/2$ with
a dominant linear dependence on $S-S_c$ in the thermodynamic limit. This
is demonstrated with the finite-$M$ scaling for different values of $S$
in Fig.~\ref{gap_AGR}(f).

In fact, for AGNRs with vertical strains, the gap opening implies a
quantum phase transition occurring at $S=S_c$ from a Fermi liquid to a
dimerized solid phase. As shown in Fig.~\ref{band_AGR}(a), $t_2$
increases and $t_{1,3}$ decrease with increasing $S$ for $\theta=\pi/2$,
which eventually leads to the dimerized $t_2$-bonds with an energy gap
$\Delta=2|t_2-2t_1|$ opening. The critical strain $S_c=0.235$ can be
determined by solving the equation $\Delta(S)=0$, or equivalently
$t_2(S)=2t_1(S)$\cite{strain-TB}. It is then understandable that the
gaps of AGNRs with finite widths diminish to zero as $S$ approaching
$S_c$ from the left, since a quantum phase transition occurs there. For
$\theta=0$, all $t_i$'s decrease with $S$ monotonically, and $t_2$
decreases even faster, hence there is no phase transition at all.

\subsection{Strained ZGNRs}
\label{sec:strained-zgnrs}
\begin{figure}[htb]
\includegraphics[width=8.6cm]{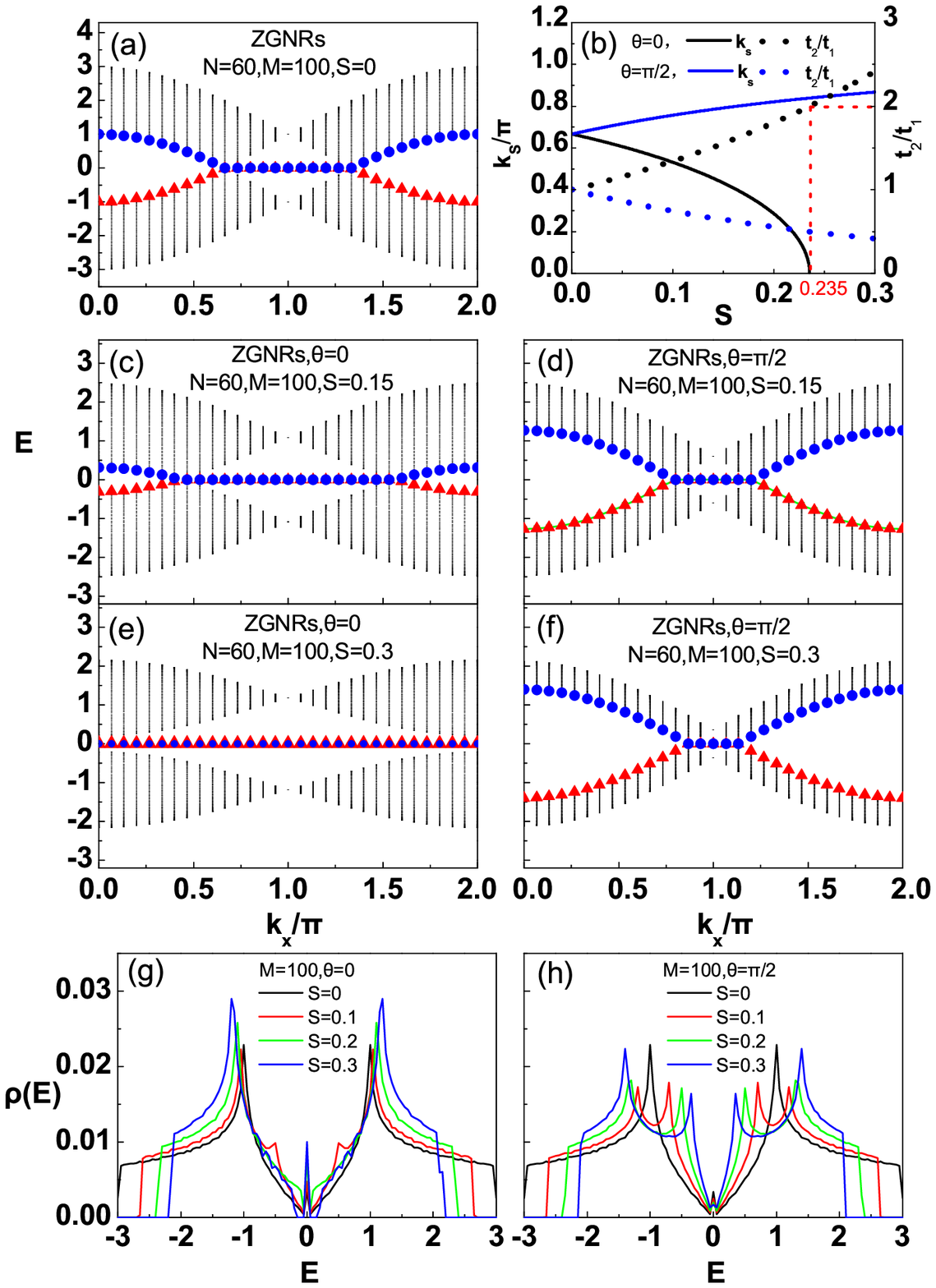}
\caption{\label{band_ZGR}(Color online.) Band structures
    of ZGNRs with $N=60$ and $M=100$ for various sizes and strains in
    (a) and (c)-(f).
  (g) and (h) are the density of states $\rho(E)$ of strained ZGNRs with
  $M=100$ and $N=1600$ for $\theta=0$ and $\theta=\pi/2$,
  respectively. (b) shows the strain dependence of $k_s$ and
  $t_2/t_1$(see text for details). }
\end{figure}
In this subsection, we discuss the band structure of ZGNRs with the
periodical boundary condition in $x$ direction and open boundary
condition in the $y$ direction. Fig.~\ref{band_ZGR}(a) is the band
structure of unstrained ZGNRs, which shows a midgap flat band
corresponding to the edge states\cite{Zigzag-EdgeState} localized in the
upper and lower zigzag boundaries. The flat band exists in a finite
region of momentum $[k_s,2\pi-k_s]$ with $k_s=2\pi/3$ for the unstrained
ribbon determined by the convergent condition for the wavefunction of
the edge states $|2\cos(k_{x}/2)|\leq1$ \cite{Zigzag-EdgeState}.

When the tension is applied, apart from those effects on the valence and
conduction bands, the region of the momentum for the flat band is also
affected by strain as seen in Fig.~\ref{band_ZGR}. For the horizontal
strain of $\theta=0$ as shown in Figs.~\ref{band_ZGR}(c) and (e), $k_s$
moves towards zero with increasing $S$ until $S=S_c$, after that $k_s=0$
and the flat band with zero energy extends over the whole Brillouin zone
accompanied by the conduction and valence bands detached from each
other. $S_c$ is the same as that defined for AGNRs in previous
subsection which signals the dimerization of the $t_2$-bonds. In
contrast, for the vertical strain of $\theta=\pi/2$, $k_s$ moves toward
$\pi$ with increasing $S$ and the flat band shrinks into a single point
with $k_s=\pi$ in the large $S$ limit as shown in
Figs.~\ref{band_ZGR}(d) and (f).

In fact, the range of the momentum for the flat band is given by the
convergent condition on the wave function, which requests
$|2t_{1}/t_{2}\cos(k_{x}/2)|\leq 1$ leading to
$k_s=2\cos^{-1}(t_2/2t_1)$ \cite{strain-band structure1}.  We plot $k_s$
and $t_2/t_1$ as functions of strain $S$ in Fig.~\ref{band_ZGR}(b). When
$\theta=0$, $t_2/(2t_1)\leq 1$ holds only for $S\leq S_c=0.235$, where
$k_s$ has a solution between $0$ and $\pi$. If $S>S_c$, $t_2/(2t_1)>1$
and the convergent condition holds for all possible momentum $k_x$,
therefore the flat band extends throughout the whole Brillouin
zone. When $\theta=\pi/2$, $t_2/(2t_1)\leq 1$ is satisfied for any
positive $S$. In this case, $k_s$ always has a solution between $0$ and
$\pi$. We note that the flat band can shrink into a point with $k_s=\pi$
if $t_2=0$, which corresponds to the horizontal $t_2$-bonds broken and
the ribbon becomes $M$ independent carbon chains connecting the left and
right electrodes.

The edge states are also revealed by the zero energy peak in the DOS
shown in Figs.~\ref{band_ZGR}(g) and (h). As the tensile strain
increases, the peak intensity is enhanced for $\theta=0$, while it is
suppressed for $\theta=\pi/2$. This coincides with previous analysis for
the region of momentum allowed for the edge states. Similar to the
AGNRs, the band width is $D=2t_1+t_2$ with a minor correction
proportional to $M^{-1}$, which also shrinks with increasing $S$. In
fact, except for the additional zero energy peaks, the characteristics
of $\rho(E)$ for ZGNRs under the uniaxial strains with $\theta=0$(or
$\theta=\pi/2$) are quite similar to those for AGNRs with
$\theta=\pi/2$(or $\theta=0$), including the positions of the shoulders
for $S=0.1$ and of double peaks, since the lattice coordinates of ZGNRs
can be obtained from that of AGNRs rotated by $\pi/2$.

\section{Transport properties of Graphene Nanoribbons}
The interplay between the strain and the finite size effect leads to the
fine tuning of the band structures of GNRs as presented in the previous
section. This allows GNRs to be considered as a promising candidate for
mechanically controllable electronic nano-devices. In this section, we use
the transfer matrix method described in Sec. II B to explore in details the
transport properties of strained GNRs with various sizes and gate voltages
as well. The transport properties essentially depend on both the band
structures of GNRs and two leads. Transport results discussed in this section
are expected to bring some insights into the designation of GNR-based
nano-devices for experimentalists.

\subsection{Conductance of strained AGNRs}
\label{sec:cond-stra-agnrs}
\begin{figure}[htb]
\includegraphics[width=8.7cm]{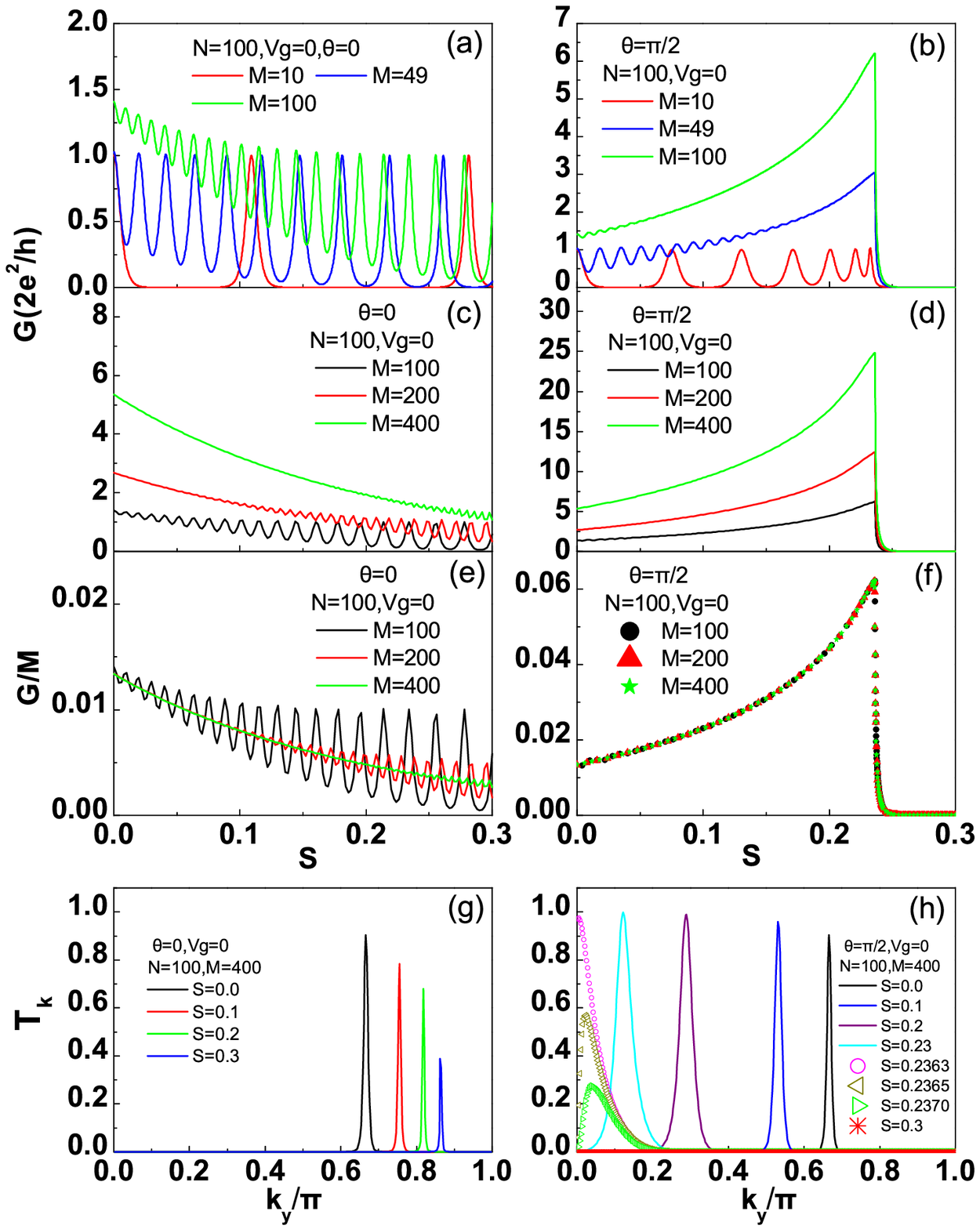}
\caption{\label{DiracPoint}(Color online.) The strain dependence of the
  conductance in charge neutral AGNRs (i.e. $E=V_g=0$) for different
  ribbon widths $M$. The tension is along $x$-axis in (a), (c) and (e),
  and along $y$ axis in (b), (d) and (f). (e) and (f) show the universal
  behavior of the scaled conductance $G/M$ with $M\geq N$ particularly.
  (g) and (h) display the transmission probabilities in different channels
  for $\theta=0$ and $\theta=\pi/2$, respectively.}
\end{figure}

We first discuss the conductance $G$ of neutral GNRs, i.e., the gate
voltage $V_g=0$. As discussed in Sec.~\ref{sec:strained-agnrs}, the band
gaps oscillate with the strains and ribbon widths, which signals one
sort of the metal-semiconductor
transition\cite{strain-transport1,strain-transport3}. This results in
that the conductances in Fig.~\ref{DiracPoint} also oscillate
accordingly, where the conductance peaks are precisely located at the
gapless points of Fig.~\ref{gap_AGR}.

Figures~\ref{DiracPoint}(a) and (b) show the conductances for the narrow
ribbons with $M\leq N$. When $M$ is small enough, say $M=10$, all the
maxima of $G$ equal $2e^2/h$, which implies that only one effective
conducting channel is maximally opened due to the strong confinement in
the $y$ direction. As the ribbon width increases from $10$ to $400$ as
shown in Figs.~\ref{DiracPoint}(a-d), more and more channels are
involved in the electronic transport, leading to the enhancement of the
conductance. In fact $G$ is almost linear in $M$ for AGNRs with fixed
lengths, which is reflected by the universal strain dependence of $G/M$
in Fig.~\ref{DiracPoint}(e,f). Besides the ribbon width, the hopping
amplitude $t_2$ also has a positive correlation with the conductance. As
one can see in an extreme situation as indicated by the geometry of AGNRs
in Fig.~\ref{QW_GRs}, that if $t_2=0$, the electronic transport would be
completely shut down. At the same time $t_2$ is controlled by the
strain, which monotonically decreases for $\theta=0$ and increases for
$\theta=\pi/2$ when $S$ increases as shown in
Fig.~\ref{band_AGR}(a). This explains the strikingly different strain
dependence of the conductance in Fig.~\ref{DiracPoint}(c) for $\theta=0$
and in Fig.~\ref{DiracPoint}(d) for $\theta=\pi/2$ (with $S<S_c$),
respectively. Figs.~\ref{DiracPoint}(a)-(f) also indicate that the
conductance oscillation is greatly suppressed in wider ribbons. For the
ribbons with the same size, the oscillation is obviously violently under
the horizontal strain than that under the vertical one.

It is interesting to note that when $\theta=\pi/2$, the conductance of
AGNRs vanishes completely in the region $S>S_c$ for any widths. This is
because a gap is opened in this region mainly by the uniaxial strain, on
which the ribbon width has little effect. In fact there is a quantum
phase transition occurring at the critical strain $S_c$ as we have
discussed in Sec.~\ref{sec:strained-agnrs}. Correspondingly, we find a
$\lambda$-like in the strain dependence of the conductance in
Figs.~\ref{DiracPoint}(d,f). The sudden drop of the conductance is
expected useful in the identification of the tension-driven phase
transition, as well as the determination of the critical strain
accurately via electronic measurements.

To further understand the electronic transport features of AGNRs, we
plot the transmission probability $T_k(k_{y,n})\equiv
\sum_{n^{'}}\eta_{nn^{'}}|t_{nn^{'}}|^2$ \cite{tm1} under different circumstances
as functions of $k_{y,n}$ in Fig.~\ref{DiracPoint}(g) for $\theta=0$ and
Fig.~\ref{DiracPoint}(h) for $\theta=\pi/2$. One can see then $T_k$ has
a spike at the momentum $k_s$, which is exactly the onset momentum of
the flat band in the spectra of ZGNRs with the same strain as seen in
Figs.~\ref{band_ZGR}. In fact, the interface between each lead and the
AGNR has a zigzag pattern, where localized states might exist similar to
the edge states in ZGNRs. As long as $k_y$ is close enough to $k_s$, the
localization length is comparable to the ribbon
length\cite{tm2,Zigzag-EdgeState,Zigzag-EdgeState2}. Therefore the
corresponding quantum states extend from the left lead to the right one,
giving the major contributions to the conductances. For $\theta=0$, the
peak position $k_s$ moves from $2\pi/3$ towards $\pi$ and the peak
height decreases with increasing $S$. However, when $\theta=\pi/2$,
$k_s$ moves towards zero and the height increases slightly as $S$
increases, until $S=S_c$. After that, the peak position shifts backward
and the height drops rapidly when $S>S_c$. For $S=0$, the analytic $T_k$
obtained in Ref.\onlinecite{tm1} gives rise to $GN/M=4e^2/3h$ at $V_g=0$
as $N,M\rightarrow\infty$ and $M/N \gg 1$. This finite value is the
maximal value for $\theta=0$ and all $S$, but the minimal value for
$\theta=\pi/2$ and $S<S_c$.

\begin{figure}[htb]
\includegraphics[width=8.7cm]{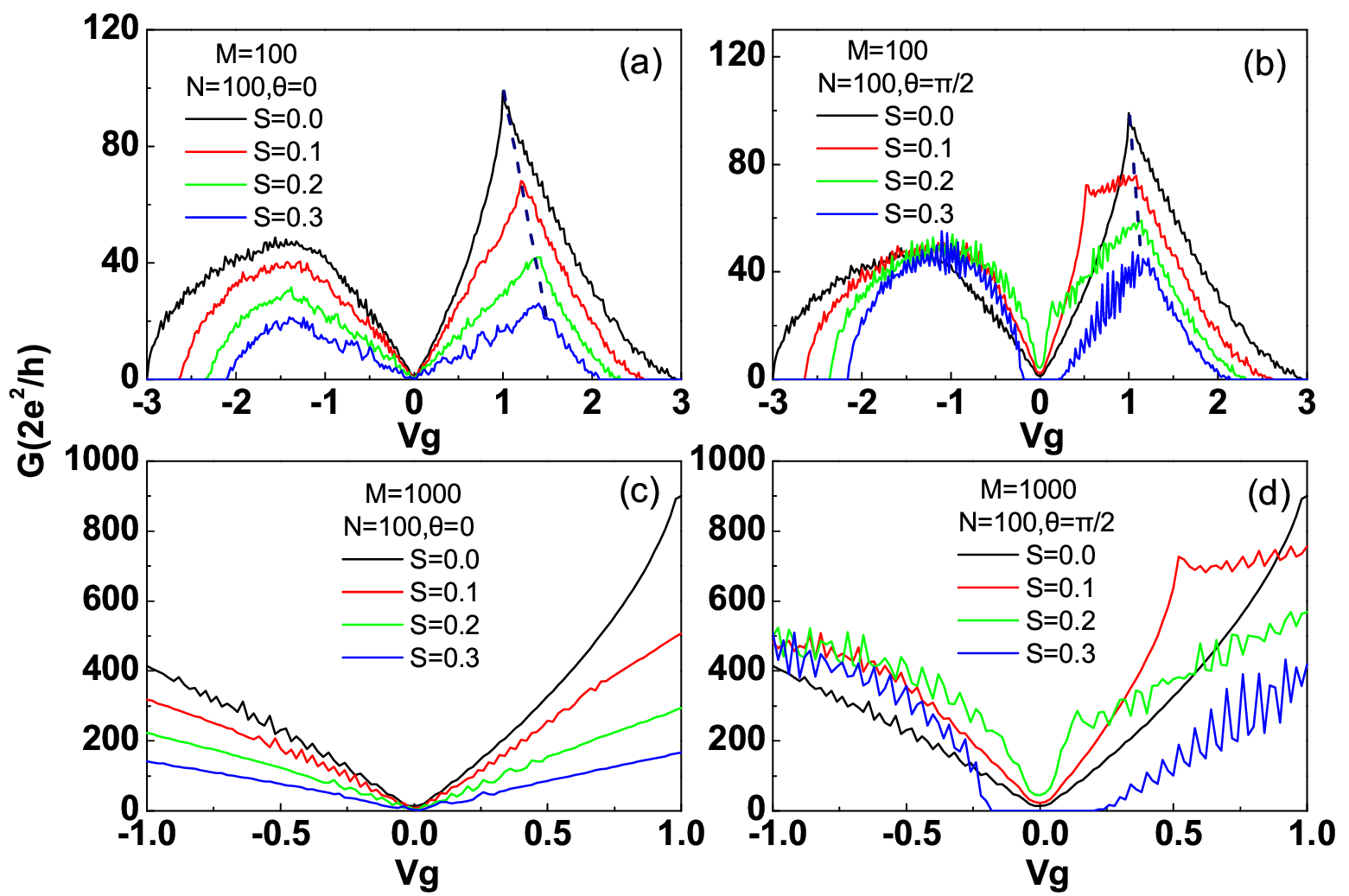}
\caption{\label{G1_AGR}(Color online.) The conductance of AGNRs as a
  function of the gate voltage $V_g$: (a) and (c) for $\theta=0$, (b)
  and (d) for $\theta=\pi/2$. The size parameters: (a) and (b) with
  $N=100$ and $M=100$, (c) and (d) with $N=100$, $M=1000$. }
\end{figure}

Figure~\ref{G1_AGR} with $M=N=100$ shows the overall features of the
conductance as a function of the gate voltage for AGNRs. One can see
that the conductance is not symmetric with respect to $V_g=0$, revealing
an electron-hole asymmetry, which has been observed in many
experiments\cite{graphene1,graphene2,graphene3-e}. This is a direct
consequence of using ordinary metallic leads\cite{tm1}. It is well-known
that a tight binding model on a bipartite lattice with only nearest
neighbor hopping is e-h symmetric. In the present system, the interface
between each lead and the AGNR consists of five-atom rings, which cannot
be bipartite and breaks the e-h symmetry eventually\cite{tm2}. One can
also see that the conductance fluctuates with $V_g$, which is due to the
scattering of electrons off the lead-ribbon interfaces and the armchair
edges, since there is no impurity and disorder in the present
system. The edges reflection of AGNRs can be suppressed relatively by
increasing the width as demonstrated in Figs.~\ref{G1_AGR}(c) and (d)
with $M=1000$, especially for small gate voltage and strains. The
remaining fluctuations in Figs.~\ref{G1_AGR}(c,d) should be attributed
essentially to the scattering on the lead-ribbon interfaces.

In Fig.~\ref{G1_AGR}(a) for $\theta=0$, the conductance curves are
rather smooth for $V_g<0$ and show a cusp at $V_g=2t_1-t_2>0$
corresponding to the higher energy peak of $\rho(E)$ in
Fig.~\ref{band_AGR}(g) and moves outwards as $S$ increases. However, the
lower energy peak of $\rho(E)$ at $E=t_2$ shows no evidence in the
conductance curves. When $\theta=0$, the conductances are suppressed by
increasing $S$ for all $V_g$ as observed in previous
studies\cite{strain-e1,strain-e2}. As a contrast, when $\theta=\pi/2$,
the conductance in Fig.~\ref{G1_AGR}(b) shows a more complicated
$V_{g}$-dependence, which reflects the significant differences between
the strain dependences of the band structures in the two cases. When $S$
increases but is still smaller than $S_c$, we find the conductance dome
in the negative energy region, an abrupt increase for $0<V_g<2t_1-t_2$,
a gentle slop for $2t_1-t_2<V_g<t_2$, and finally a decreasing region
for $V_g>t_2$. Although the hopping amplitudes $t_i$'s shift with $S$,
we can still claim that, $G$ is an increasing function of $S$ for small
$V_g$, and a decreasing function for large $V_g$.  When $S\geq S_c$, a
gap $\Delta$ opens and $G$ vanishes for $V_{g}<\Delta$ and is suppressed
by increasing $S$ for any $V_g>\Delta$ as can be seen in
Figs.~\ref{G1_AGR}(b) and (d).

\begin{figure}[htb]
\includegraphics[width=8.7cm]{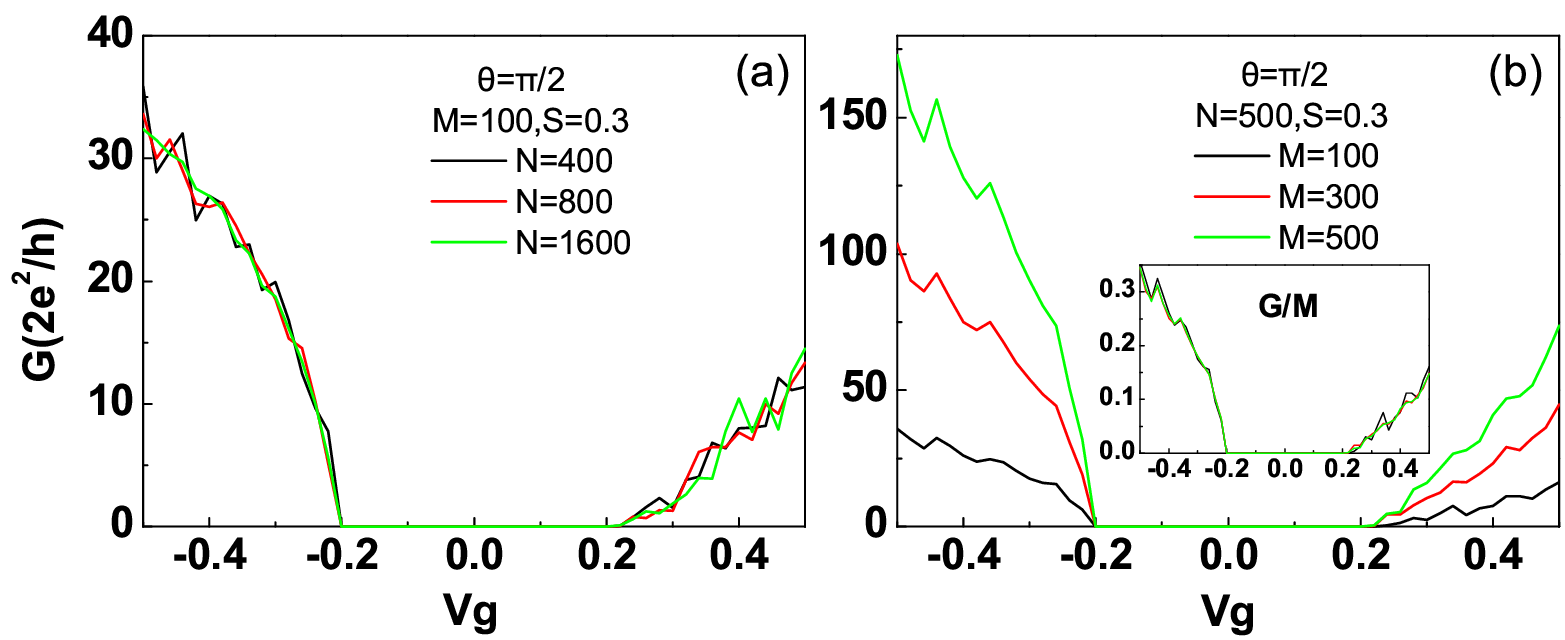}
\caption{\label{G2_AGR}(Color online.) The conductance of AGNRs for the vertical strain $S=0.3>S_c$ with
  various lengths $N$ and widths $M$. Inset in (b): the conductance scaled by the width $M$. }
\end{figure}

Figures~\ref{G2_AGR}(a) and (b) show more details on $G$ for $S=0.3$
and $-0.5t_0<V_g<0.5t_0$. The conductance is zero for $-0.2t_0< V_g<
0.2t_0$, which indicates $\Delta=0.4t_0$ for $S=0.3$ in consistency with
that from the direct calculation given in Fig.~\ref{gap_AGR}(e).
Fig.~\ref{G2_AGR}(a) also implies that as $N\ge400$ the conductance
barely changes with increasing $N$ due to the ballistic
transport. Fig.~\ref{G2_AGR}(b) shows the dependence of the conductance
on $M$ and its inset reveals the universal behavior of renormalized
conductance $G/M$.

\subsection{Conductance of strained ZGNRs}
\label{sec:cond-stra-zgnrs}
\begin{figure}[htb]
\includegraphics[width=8.7cm]{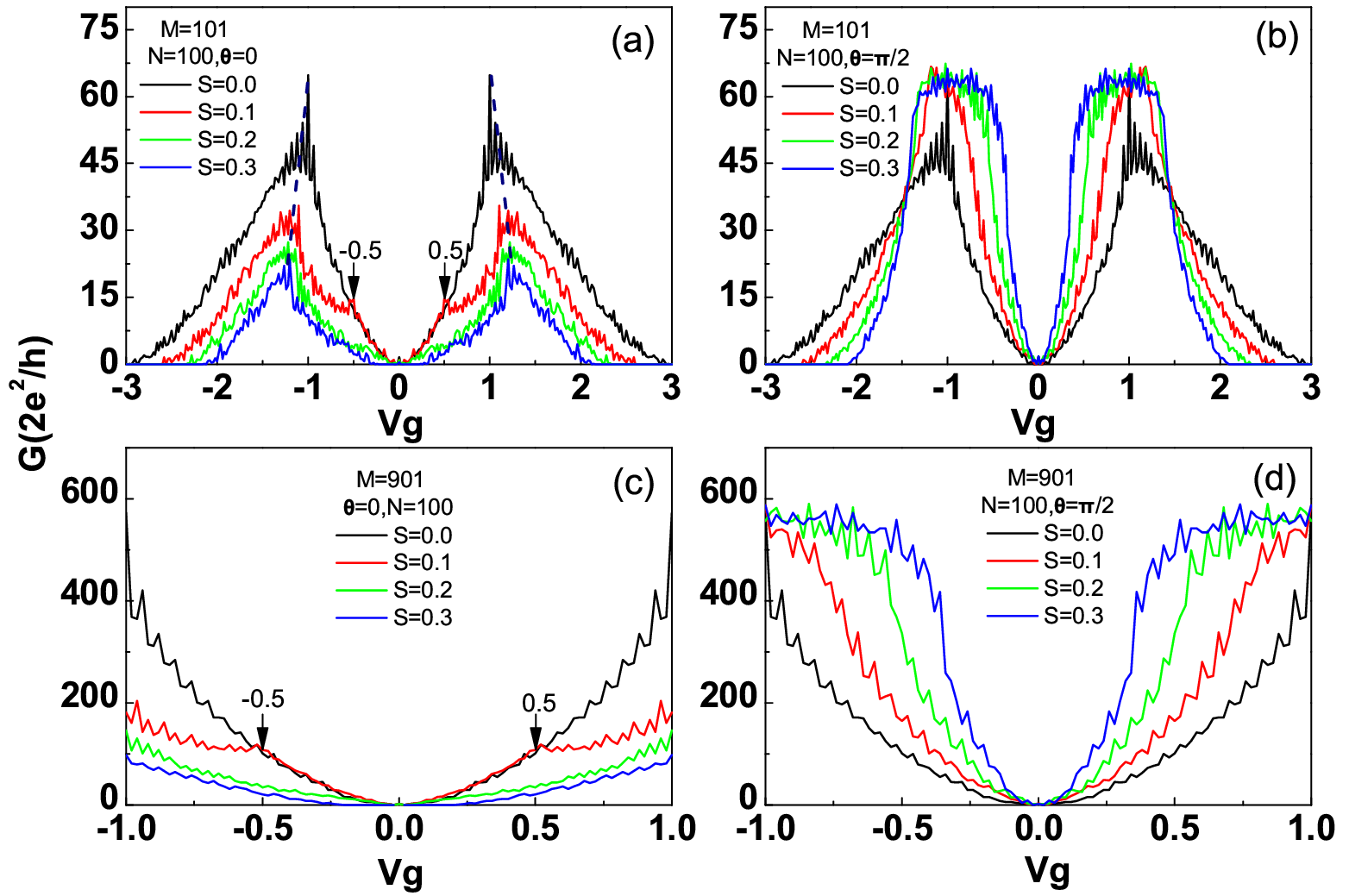}
\caption{\label{G2 ZGR} (Color online.) The conductance of ZGNRs as a
  function of the gate voltage $V_g$: (a) and (c) for $\theta=0$; (b)
  and (d) for $\theta=\pi/2$. The size parameters: (a) and (b) for
  $N=100$ and $M=101$; (c) and (d) for $N=100$ and $M=901$. }
\end{figure}

Figures~\ref{G2 ZGR}(a)-(d) display the conductance of the ZGNRs as a
function of $V_g$ with various $S$ and $M$ for fixed $N=100$. In
particular, Figs. \ref{G2 ZGR}(a) and (b) with $M=101$ show the overall
features of $G$ in the full range of the band width, while Figs.~\ref{G2
  ZGR}(c) and (d) with a larger $M=901$ is shown to demonstrate less
conductance fluctuations for $|V_g|\leq 1.0$. The conductance data
obviously shows the e-h symmetry unlike the AGNRs case. This is because
the whole system is still bipartite since those rings on the interfaces
between the leads and ZGNR contain either four or six atoms as seen in
Fig.~\ref{QW_GRs}(b), in contrast to the non-bipartite five-atom rings
on the interfaces in the system of AGNRs. It thus seems that all the
features of the conductance are essentially consistent with the DOS for
ZGNRs in Fig.~\ref{band_ZGR}. It is also remarkable that the conductance
is a constant around $G_0$($G_0\equiv
2e^2/h $) or vanishingly small at zero gate voltage in
the large $N$ and $M$ limit and the flat band is apparently not involved
in electronic transport even at $V_g=0$.  This feature is unchanged
under the strains and we will discuss it latter.

In Fig.~\ref{G2 ZGR}(a) for $\theta=0$, two sharp peaks of the
conductance for each different strain correspond to the peaks of
$\rho(E)$ at $E=\pm t_2$ in Fig.~\ref{band_ZGR}(g), which move outwards
when the strain increases. It is interesting to note that in Figs.~\ref{G2
  ZGR}(a) and (c) the conductance for $S=0.1$ is almost identical to
that for $S=0$ in the region $|V_g|<0.5t_0$. In fact this phenomena
emerges for any given strain $S<S_c=0.235$, and the overlapping region
between $G(S)$ and $G(0)$ is given by $|V_g|< 2t_1-t_2$ which is
obviously strain dependent. In other word, given a small $V_g$, the
relation $|V_g|=2t_1-t_2$ gives rise to a threshold of strain, below
which the measured conductance barely changes with respect to
$S$. Despite of this identical region, Fig.~\ref{G2 ZGR}(a) for
$\theta=0$ also indicates that the conductance is reduced with
increasing $S$ if the gate voltage is fixed. Figs.~\ref{G2 ZGR}(b) and
(d) show the conductance for $\theta=\pi/2$, where we find a cross point
of the conductance curves under different strains. For convenience we
denote the corresponding gate voltage as $V_g^C$ which is around
$1.5t_0$. The strain enhances the conductance for $|V_g|< V_g^C$ and
suppresses it otherwise.

To interpret the difference of the conductances between $\theta=0$ and
$\theta=\pi/2$, we recall that the hopping integrals $t_i$'s have
different strain dependence as seen in Fig. \ref{band_AGR}(a). When the
strain increases for $\theta=0$, $t_2$ increases, while $t_{1,3}$
decreases. However, $t_{1,3}$ effectively favors the horizontal
electronic transport, while $t_2$ may cause the formation of the dimers
for vertical bonds which hinders the electrons from moving
freely. Therefore, the conductance is reduced by increasing the strain
for given $V_g$. However, for $\theta=\pi/2$, both $t_{1,3}$ and $t_2$
decrease, but $t_2$ drops faster. As a consequence, the ribbon tends to
form $M$ metallic chains with a weak interchain coupling. The
conductance is then enhanced with an upper limit $MG_0$ as $t_2\rightarrow0$ for small $V_g$. Since more and more
channels below $V_g^{C}$ are fully filled due to the reduction of the
hopping amplitudes $t_{1,3}$, they do not contribute to the conductance
for large gate voltage.  It turns out that there are two turning points
$\pm V_g^C$ in Fig.~\ref{G2 ZGR}(b) and opposite strain dependences of
the conductance are found for $|V_g|<V_g^C$ and for $|V_g|>V_g^C$,
respectively.

\begin{figure}[htb]
\includegraphics[width=8.cm]{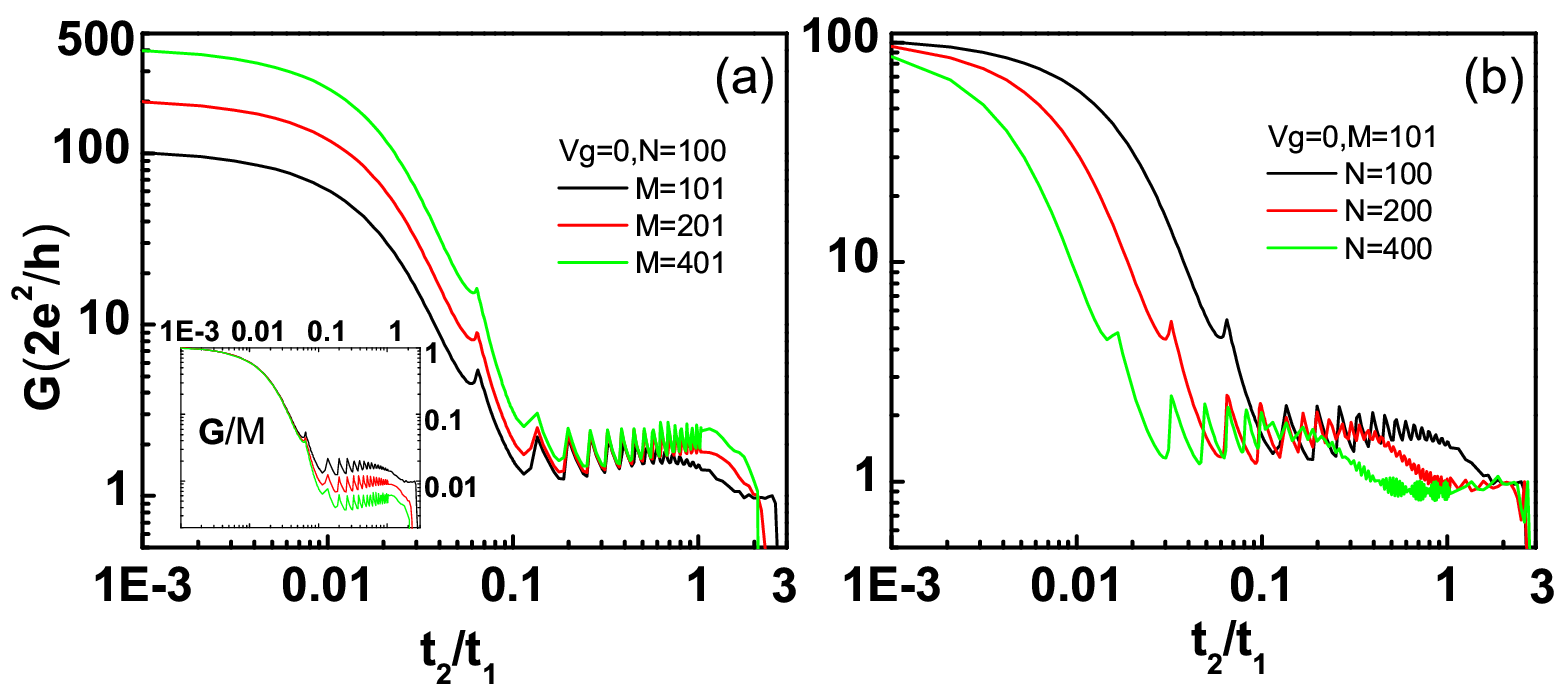}
\caption{\label{t2_ZGR} (Color online.) The conductance versus $t_2/t_1$ at $V_{g}=0$ with
  $t_1=0.96$ fixed. (a) for $N=100$ and $M=101$, 201 and 401; and (b) for $M=101$ and $N=100$, 200 and 400.  Inset in (a) for the conductance scaled by the width $M$.}
\end{figure}

For ZGNRs, the topology of structure not only protect the e-h symmetry
of the electronic transport but also stimulates the analysis of
more general features of the conductance in a whole range of
$t_2/t_1$, which might be beyond the values given by the relations
Eqs. (\ref{eq:3}). In principle, one has actually two limits:
$t_2/t_1\gg 1$ and $t_2/t_1\ll 1$. For the former case, the system
possesses an ordered and insulating ground state consisting of dimerized
$t_2$-bonds, which already emerges actually with a zero conductance at
$t_2/t_1\approx 2.9$ for $M=101$ and $N\gtrsim M$ as demonstrated in
Fig. \ref{t2_ZGR}(a) and (b). When $t_2/t_1\ll 1$, the honeycomb lattice
becomes $M$ weakly coupled metallic (zigzag) chains and the conductance
reaches its maximal value $MG_0$ as seen in Fig.~\ref{t2_ZGR}(a) where
the conductance is shown as a function of $t_2/t_1$ for $N=100$ with
$M=101,201$ and $401$. One can see that $G$ is enhanced by decreasing
$t_2/t_1$ and indeed proportional to $M$ as being well renormalized by
$M$ for $t_2/t_1\lesssim 0.05$ in the inset.  Fig. \ref{t2_ZGR}(b) shows
that the conductance is also reduced by increasing the length of
ribbons, implying a non-ballistic transport. In addition, the
conductance fluctuations around $G_{0}$ show up for $t_2/t_1\gtrsim
0.1$.

\begin{figure}[htb]
\includegraphics[width=8.0cm]{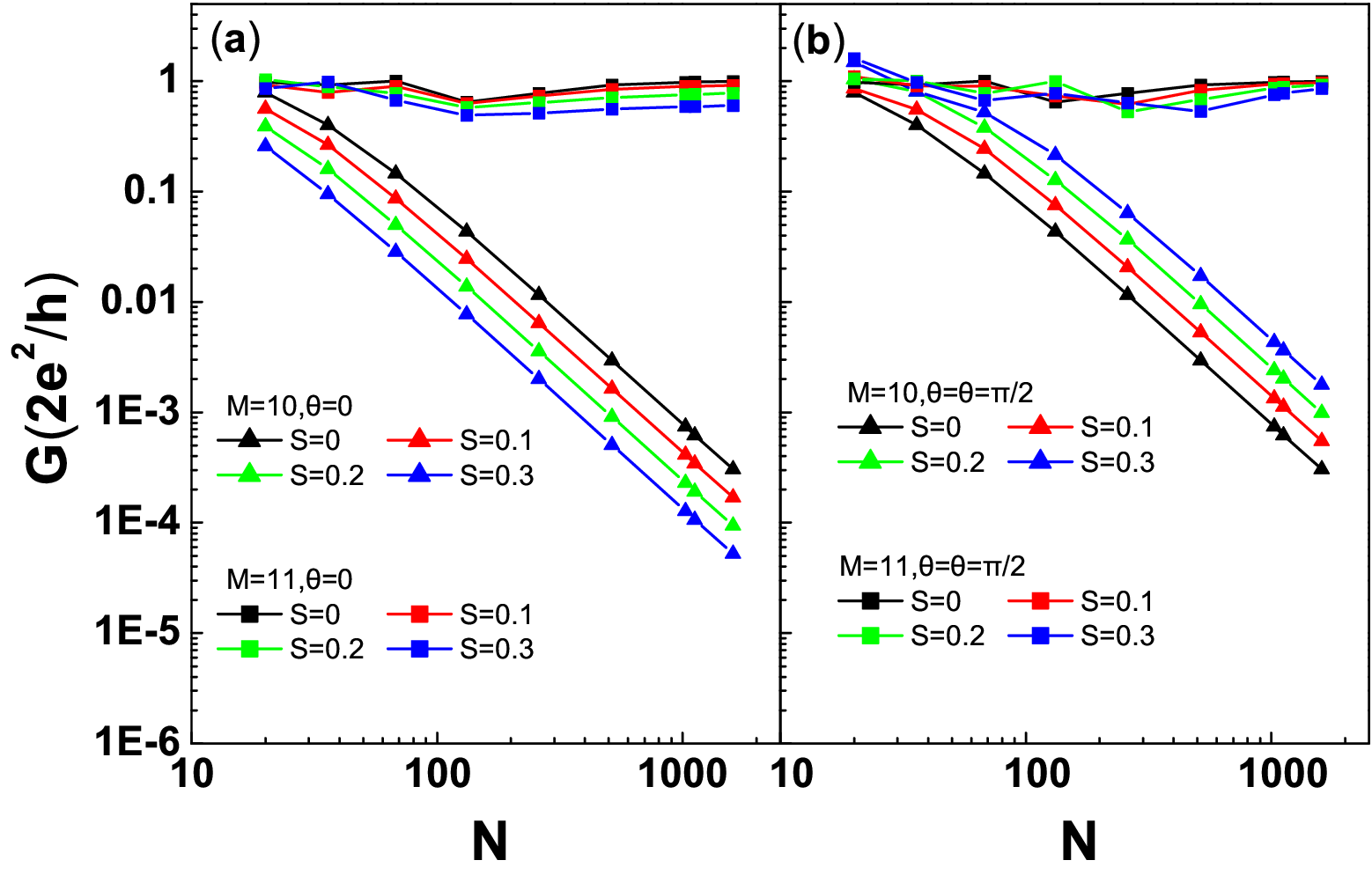}
\caption{\label{DiracPoint-Z}(Color online.) Odd-even
    $M$ effects on the conductance for narrow ZGNRs for $\theta=0$ in
    (a) and $\theta=\pi/2$ in (b).}
\end{figure}
Figure~\ref{DiracPoint-Z} shows an even-odd $M$ effect on $G$ for ZGNRs,
which is relevant for either experiments or designing nano-devices with
narrow and short ribbons. This effect diminishes for sufficiently wide and
long ribbons so that our above discussions for Figs. \ref{G2 ZGR} and
\ref{t2_ZGR} are given just for odd but sufficiently large $M$. For
$S=0$ and $V_g=0$, one finds that the conductance of narrow ZGNRs shows
two different scaling behaviors according to the parity of $M$ as
$N\rightarrow\infty$\cite{ZGNR-conductance,ZGNR-conductance2}. The
conductance is a constant around $G_0$ for odd $M$ to indicate metallic
nature, while $G\sim N^{-2}$ for even $M$ to present a semiconducting
feature \cite{tm2}. In the presence of strains, one can still find
two types of scaling behaviors for the conductance at $V_g=0$ as shown
in Figs. \ref{DiracPoint-Z}(a) and (b). When $M$ is odd, the conductance
changes a little with both $N$ and $S$. When $M$ is even, although the
conductance decreases in a power law $N^{-2}$, $G$ is suppressed for
$\theta=0$ and enhanced for $\theta=\pi/2$ with increasing $S$ if $N$ is
fixed.

\section{Effects of edge relaxation}

  In realistic situations, there may exist passivation and
  spin poarlization on the ribbon edges, which affect the band structure
  of narrow ribbons \cite{band structure-AGNR}.  However, these edge modes
  only affect the edge and nearby carbon atoms for wide enough ribbons,
  in contrast, the strain can affect all the carbon-carbon bonds in the
  ribbon. As shown in Fig.3, the band gaps oscillate with the strain for all types of AGNRs and is around $0.25t_0$ for $M=10$ (i.e., $N_a=20$), while the edge relaxation induces one only around $0.032t_0$ as shown in Ref.28. This indicates the edge relaxation has much smaller effect on the band gap than the strain effect, which even becomes smaller and smaller as the ribbon width increases.

\begin{figure}[htb]
\includegraphics[width=8.0cm]{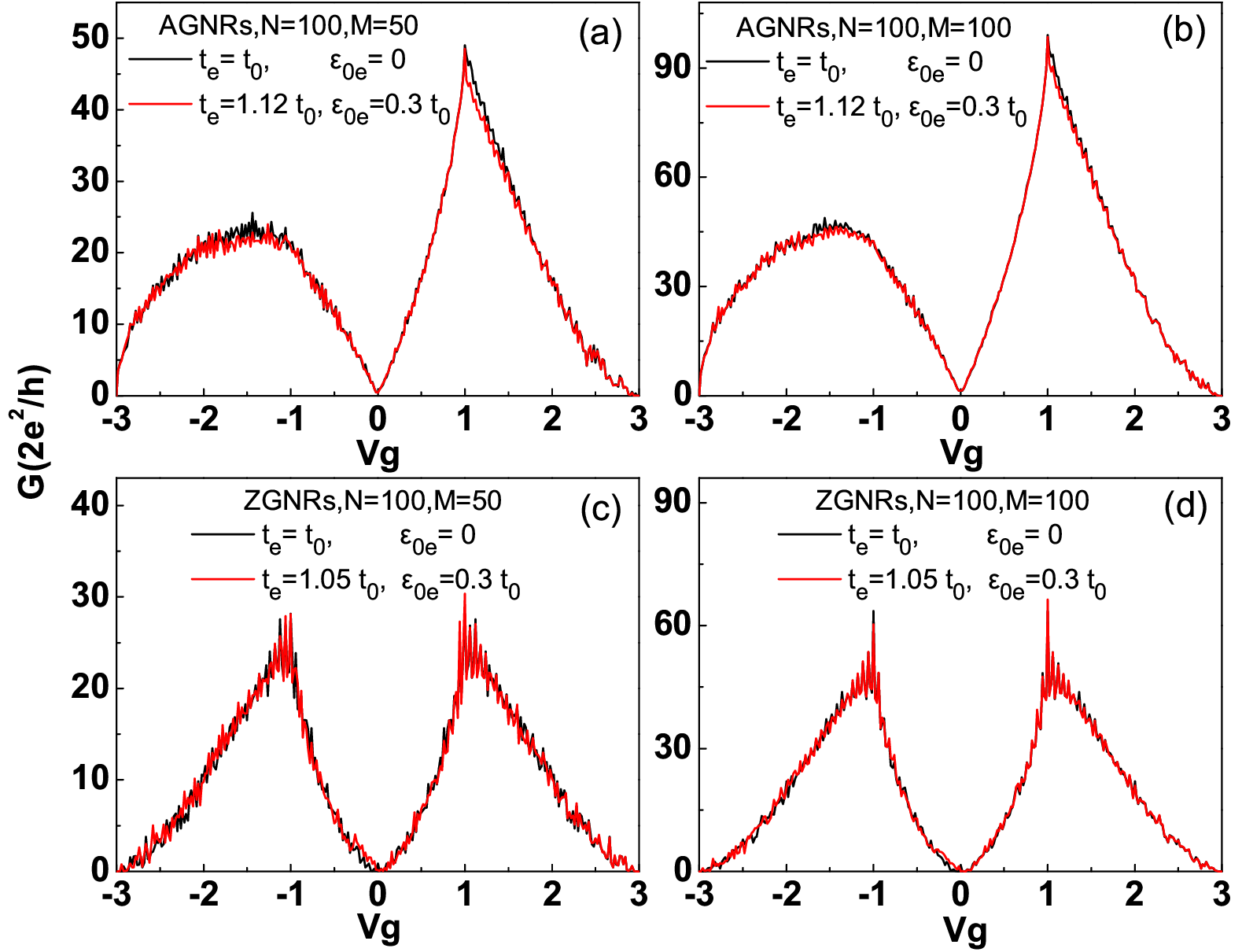}
\caption{\label{edge-effect}(Color online.) Effects of edge relaxation on the conductance of AGNRs (a,b) and ZGNRs (c,d) with various sizes.}
\end{figure}

Comparing with the strain effect, edge relaxation on the transport properties of GNRs is negligible in relatively wide graphene ribbons. In Fig. \ref{edge-effect}, the black solid lines are the conductance of pristine GNRs and the others are the data with edge relaxation. For AGNRs, we set the hopping integral at the edge te to be $1.12t_0$ as suggested in Ref. 28 and the on-site energies $\epsilon_{0e}$ being $0.1t_0$ and $0.2t_0$. It is clearly seen that the effects of edge relaxations on electronic transport are negligible as anticipated for armchair ribbons wider than 10nm (i.e., $M\geq42$). For ZGNRs, it is well known that tight binding parameters of carbon atoms based on first-principle calculation are environment dependent, such as QUAMBOs tight binding (QUAMBOs-TB) parameters as shown in \cite{edge-effect}. We set the hopping integral at the edge $t_e=1.05t_0$ and the on-site energies $\epsilon_{0e}$ being $0.25t_0$ according to \cite{edge-effect}. It is clearly seen that the effects of edge relaxations on electronic transport are negligible for zigzag ribbons wider than 10nm (i.e., $M\geq48$).

   In simple tight binding model, edge spin polarized states of ZGNRs requires the introduction of Coulomb interaction into the flat band. However, these magnetic moments at the edges of zigzag nanoribbons (for correlation effects) are far beyond the scope of the present investigation. Similar to edge relaxation, edge spin polarization only affects the edge atoms and these spin polarized edge modes can provide at most two conducting channels. Therefore, their effect on the electronic transport is expected to be less important than the strain effect which is on all the bulk conducting channels.  As a consequence, when investigating the strain effect on the transport through GNRs, one can ignore these edge effects.

\section{Summary and conclusions}
In this article, we have investigated the electronic transport of
graphene nanoribbons under various tensile strains with connections to
the normal metallic leads.  For this purpose, we first calculated the
band structures of strained GNRs with both zigzag and armchair
edges. The direction of the uniaxial tension, which is taken to be
either parallel($\theta=0$) or perpendicular($\theta=\pi/2$) to the
ribbon axis, has a crucial effect on the band structure.

In the strained armchair GNRs with $\theta=0$, the band gap oscillates
with the strain in a zigzag pattern, leading to the transitions between
metal and semiconductor. The oscillatory amplitude is almost unchanged
as $S$ increases. This kind of band gap is mainly a finite width effect,
since it vanishes as $M$ goes to infinity.  If $\theta=\pi/2$, similar
oscillatory gap also appears, but only for the strains smaller than a
critical value $S_c$. As $S$ approaches $S_c$, the oscillatory amplitude
goes to zero, unlike the case for $\theta=0$. Once $S>S_c$, the other
kind of band gap opens which is linear in $S-S_c$ and hardly affected by
the ribbon width. In fact as the strain with $\theta=\pi/2$ increases, a
quantum phase transition is induced at $S=S_c$ to separate a liquid
phase from a solid phase where the bonds perpendicular to the strain are
dimerized.

In the zigzag GNRs, the most intriguing phenomenon is the appearance of
the flat band in a region of momentum $[k_s,2\pi-k_s]$. As the strain
with $\theta=0$ increases, $k_s$ decreases to zero until $S=S_c$, then
the flat band extends throughout the full Brillouin zone, and the
conduction and valence bands are separated. On the contrary, with
increasing $S$, $k_s$ moves towards $\pi$ and the region of the flat
band shrinks into a point for $\theta=\pi/2$.

Except for the flat band, most features on the strain-dependence of the
band structures are well revealed by the behaviors of the conductance of
GNRs. For example, the band gap oscillation results in the conductance
oscillation at the zero gate voltage $V_g=0$ as the strain varies. The
peak in the plot of the conductance versus $V_g$ is compatible with that
in the DOS plot. Note that not all the modes with energy $V_g$
contribute to the conductance, but only those satisfying the boundary
conditions are responsible for the electronic transport, therefore, it
is not necessary to have a one-to-one correspondence between the peaks
of the conductance and those of the DOS.  Furthermore, by measuring
the strain dependence of the conductance of AGNRs at $V_g=0$, one can
also detect the quantum phase transition induced by the tension
perpendicular to a C-C bond and determine the critical strain as well.

Since we connect the GNRs with square lattices as the metallic
electrodes, it is worth mentioning the fundamental effect of the
topology of the heterojunctions on the conductance of GNRs.  In
particular, due to the non-bipartite feature of the electrode-AGNR
interfaces, the conductance data of AGNRs is not e-h symmetric, while
this kind of symmetry can still be found in that of ZGNRs, since the
electrode-ZGNR interfaces do not break the bipartite structure of the
whole system.  This phenomenon has no counterpart in the band structures
obtained with periodic boundary condition, yet it may be important for
designing the nano-size devices.

\section{Acknowledgements} This work is supported by the National Basic
Research Program of China (2012CB921704) and NSF of China (Grant
Nos. 11174363, 11204372, 11374135). We thank G.H. Ding for helpful
discussions.

\end{document}